\documentclass[12pt,preprint]{aastex}

\usepackage{emulateapj5}
\usepackage{onecolfloat}
\usepackage{graphicx}
\usepackage{times}

\newcommand{\sm}{\, {\rm M}_{\odot}}

\newcommand{\ndeg}{^{\rm o}}
\newcommand{\Aq}[1]{\texttt{Aq-#1}}
\def\gsim { \lower .75ex \hbox{$\sim$} \llap{\raise .27ex \hbox{$>$}} }
\def\lsim { \lower .75ex \hbox{$\sim$} \llap{\raise .27ex \hbox{$<$}} }

\newcommand{\cutt}[1]{}

\shorttitle{Substructure in the Aquarius stellar halos} 
\shortauthors{A. Helmi et al.}

\received{2011 January 11}
\accepted{2011 April 8}
\begin{document}

\twocolumn[

\title{Substructure in the stellar halos of the Aquarius simulations}

\author{Amina Helmi\altaffilmark{1}, A.~P.~Cooper\altaffilmark{2,3},
S.D.M. White\altaffilmark{3}, S. Cole\altaffilmark{2},
C.S. Frenk\altaffilmark{2}, J.F. Navarro\altaffilmark{4}}

\begin{abstract}

  We characterize substructure in the simulated stellar halos of
  Cooper et al. (2010) which were formed by the disruption of
  satellite galaxies within the cosmological N-body simulations of
  galactic halos of the
  Aquarius Project.  These stellar halos exhibit a wealth of tidal
  features: broad overdensities and very narrow faint streams akin to
  those observed around the Milky Way. The substructures are
  distributed anisotropically on the sky, a characteristic that should
  become apparent in the next generation of photometric surveys.  The
  normalized {\it RMS} of the density of stars on the sky appears to
  be systematically larger for our halos compared to the value
  estimated for the Milky Way from main sequence turn-off stars in the
  Sloan Digital Sky Survey. We show that this is likely to be due in
  part to contamination by faint QSOs and redder main sequence stars,
  and might suggest that $\sim 10\%$ of the Milky Way halo stars have
  formed in-situ.

\end{abstract}

] 

\altaffiltext{1}{Kapteyn Astronomical Institute, University of Groningen,
P.O.Box 800, 9700 AV Groningen, The Netherlands.
{\sf{e-mail: ahelmi@astro.rug.nl}}}
\altaffiltext{2}{Institute for Computational Cosmology, Department of
Physics, University of Durham, South Road, Durham DH1 3LE, UK}
\altaffiltext{3}{Max-Planck-Institut f\"{u}r Astrophysik,
Karl-Schwarzschild-Str. 1, D-85748, Garching, Germany}
\altaffiltext{4}{Department of Physics and Astronomy, University of Victoria,
  Victoria, BC V8P 5C2, Canada}

\section{Introduction}
\label{sec:intro}

Stellar halos are repositories of merger debris and hence,
despite their low luminosity, are central to efforts to unravel the
accretion history of galaxies \citep{sz}. 
Wide-field photometric surveys have discovered a plethora of
substructures in the halo of our Galaxy
\citep[e.g.][]{majewski,vasily-fos,juric}. Some examples are the Orphan Stream
\citep{vasily-os} and the Sagittarius dwarf \citep{ibata94} and its extended
tidal tails \citep{ivezic,yanny}. Other conspicuous structures, like
the Hercules-Aquila cloud \citep{vasily-ha}, the Virgo overdensity and
the Virgo Stellar Structure, do not have the elongated appearance
typical of tidal features and hence their nature is less
obvious. Moreover, some
of these structures strongly overlap on the sky but their relationship
is not always clear \citep{Virgo,VSS,newberg2007}.

The amount of spatial substructure present in the outer halo has been
quantified by \citet[][see also de Jong et al. 2010]{bell}. These authors found {\it RMS} fluctuations
of order 30--40\% with respect to a smooth halo model. Taken at face
value, this implies that at least this fraction of the stellar halo
has been accreted. \cite{else} determined that at
least 10\% of the outer halo must have been accreted, based on the
amount of clustering present in the Spaghetti Survey (which also
included radial velocity information). Roughly half of this survey's
pencil beams pointed to portions of the sky unexplored by SDSS;
intriguingly, these authors found no evidence of substructure in those
regions, suggesting that the distribution of outer halo substructure
may be very anisotropic.

Several studies have examined the properties of stellar halos in
$\Lambda$CDM simulations of structure formation. For example,
\citet{bj05} followed the accretion of satellites onto a smooth,
time-dependent gravitational potential resembling a Milky Way
galaxy. Their simulations produce very lumpy stellar halos, with
extended tidal streams covering large portions of the
sky. \cite{gdl-ah} used high-resolution cosmological N-body
simulations of the formation of a Milky-Way size dark halo in
combination with a semi-analytic model of galaxy formation. Although
their model reproduces the global properties of the Galactic stellar halo,
the resolution of the simulations was too low to show much substructure
in the form of tidal tails.

Recently, \citet[][hereafter C10]{cooper} have combined full
cosmological dark matter simulations from the Aquarius project with
the Durham semi-analytic galaxy formation model, {\sc galform}, in
order to follow the assembly of `accreted' stellar halos. The
extremely high resolution of the Aquarius halos permits a wealth of
substructure to be studied in the \citeauthor{cooper} models. This
{\it Letter} will focus on the properties of this substructure as may
be revealed by ongoing (e.g. SDSS) and future wide-field photometric
surveys, such as Skymapper, Pan-STARRS or LSST.

\section{Brief description of the model}

The Aquarius project consists of simulations of the formation of
several dark matter halos with masses comparable to that of the Milky
Way in a $\Lambda$CDM universe\footnote{Here we focus on the analysis
  of five of the six Aquarius halos, because the remaining object's
  entire stellar halo was built in a merger at $z \sim 0.3$ and is
  thus unlikely to be representative of the Milky Way.}.  The Aquarius
halos (labelled \Aq{A} to \Aq{E}) are resolved with more than $10^8$
particles, and their properties are described in
\citet{springel-sub}. To follow the evolution of the baryonic
component, the dynamical information obtained from the simulations is
coupled to the Durham semi-analytic galaxy formation model
\citep{bower}. In essence, C10 identify at each output the halos that
host galaxies according to the semi-analytic model, and tag the 1\%
most bound particles at that time with ``stellar properties'' such as
masses, luminosities, ages and metallicities. This allows us to
follow the assembly and dynamics of the accreted component of
galaxies.

Note, however, that we do not follow stellar populations formed
in-situ (for example, a disc component) because our dynamical
simulations are purely collisionless (hydrodynamics and star formation
are treated semi-analytically). A massive disc may influence satellite
orbits and their debris, and also provide a reservoir from which stars
can be scattered into the halo, but these effects are absent in our
models.

The tagging scheme leads to the identification of more than
$4\times10^5$ tracer dark matter particles for \Aq{A} and up to a
maximum of $6 \times 10^5$ for \Aq{B} and \Aq{D}.  Since the stellar
mass assigned to each of these dark matter particles depends on the
mass of stars formed in the parent galaxy with time, they
do not all carry the same weight (for more details, see
C10). Therefore, and for ease of comparison to observations, we have
resampled the tagged particles as follows. For each tagged dark matter
particle we find the 32 neighbors ranked as nearest simultaneously in
position and in velocity at the present time amongst the class of
tagged particles that share the same progenitor. We then measure the
principal axes of the spatial distribution of these neighboring
particles. This is used to define the characteristics of a
multivariate Gaussian from which we draw positions for re-sampled
stars.  We then down-sample this new set of ``stellar particles'' to
represent main sequence turn-off stars \citep[assuming $5 \sm$ per
MSTO, as used in][]{bell} or red giant branch stars \citep[we assume
1 RGB per  8 MSTO stars, as in the models
of][]{marigo}. This results in $\sim 5.5 \times 10^7$ MSTO stars
between 1 kpc and 50 kpc from the halo center for \Aq{A}, and up to $\sim
1.83 \times 10^8$ MSTO stars for \Aq{D} (which is the brightest of our
stellar halos).

\label{sec:model}
\begin{figure*}[t]
\begin{center}
\includegraphics[height=17.5cm,angle=270]{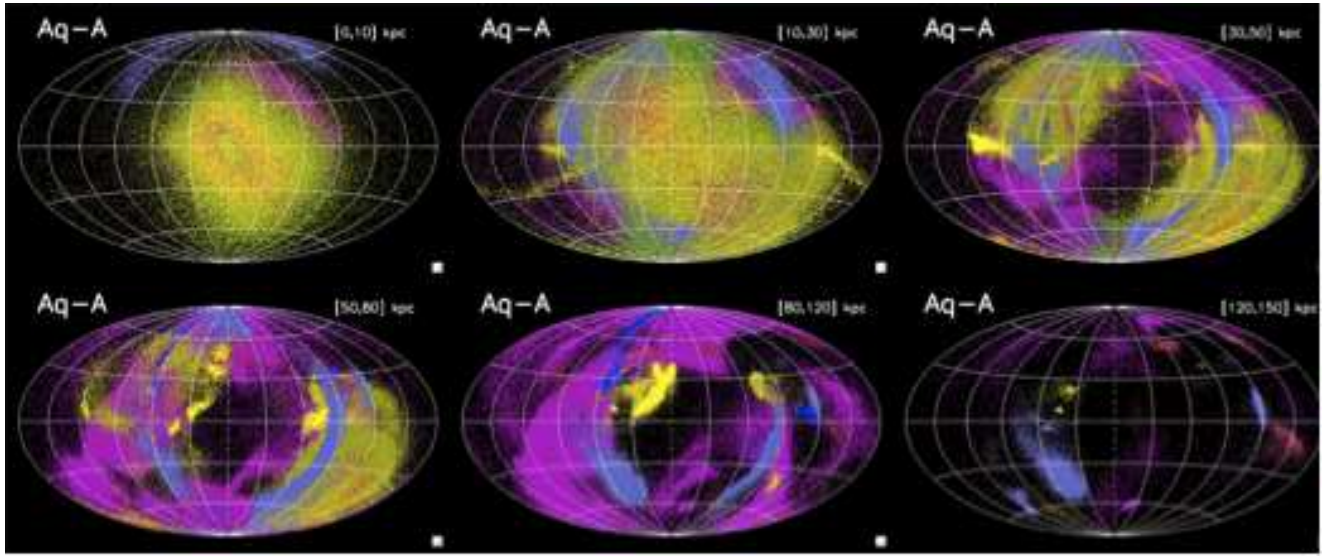}
\end{center}
\caption{Distribution of ``field'' RGB stars on the sky at various distances
from the ``Sun'' for the stellar halo of \Aq{A}. The different colors
correspond to stars originating in different progenitors. The total
number of progenitors is 163, out of which 53 contribute more than
$10^4 \sm$ in stars. There are 35 different progenitors contributing
at least 50 RGB stars in the innermost bin, and 48 at distances
between 10 and 30 kpc. This number drops down to 10 in the most
distant bin considered here. Thus the \Aq{A} stellar halo
is most diverse between 10 and 30 kpc.
 \label{fig:sky_haloA}}
\end{figure*}
\begin{figure*}
\begin{center}
\includegraphics[height=17.5cm,angle=270,clip]{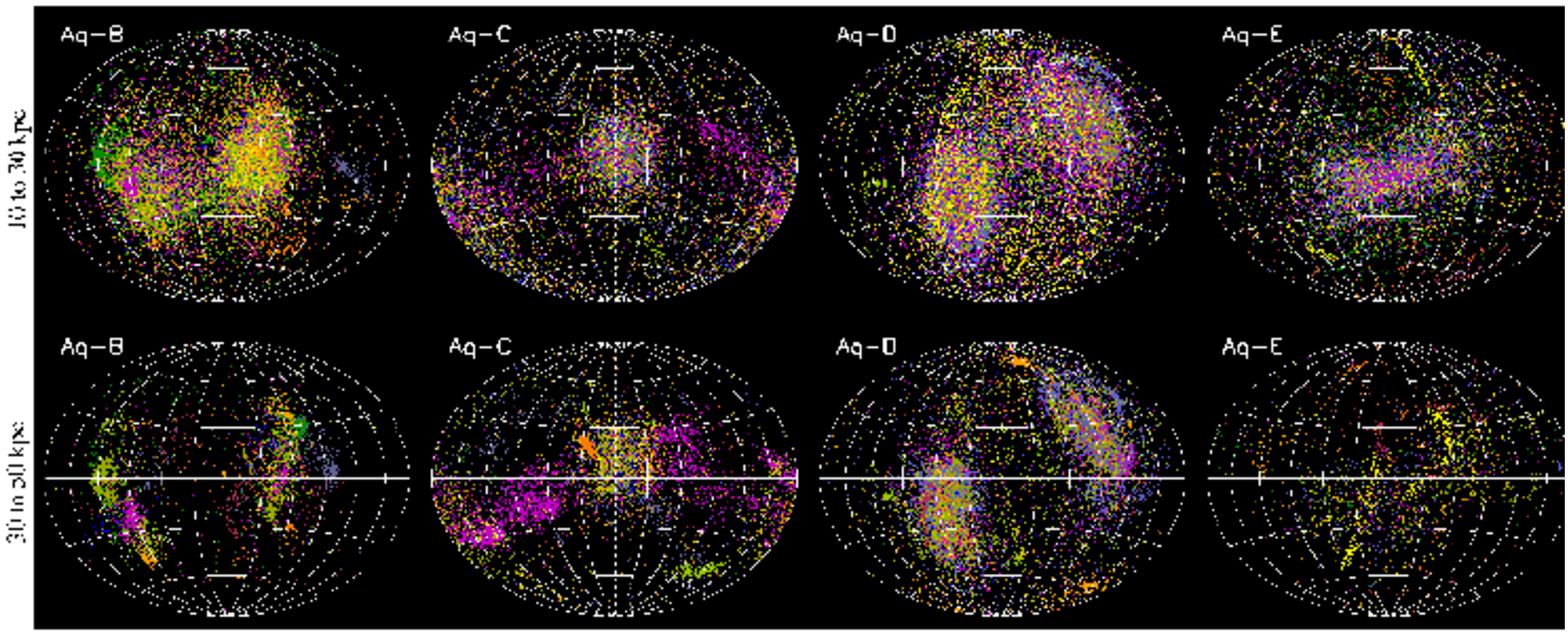}
\end{center}
\caption{Sky distribution of RGB stars located at ``heliocentric'' distances [10, 30]
  kpc (top) and [30,50] kpc (bottom) for the set of Aquarius stellar halos.
\label{fig:sky_halo_all}}
\end{figure*}

\section{Results}

In this {\it Letter} we focus on the distribution of halo stars on the sky for
direct comparison to photometric surveys. We will show that panoramic
views resembling the ``Field of Streams'' are common in our simulations,
and we will provide a new theoretical perspective on the nature of the
overdensities discovered by SDSS. At the end, we quantify 
the amount of substructure present in our stellar
halos and compare it to the results of \cite{bell}. In future papers we
will focus on the characterization of substructure (width, age,
relation to progenitor system, etc.), and on the amount of
kinematic substructure  present in the Solar neighborhood (especially
for comparison with RAVE, and Gaia data in the future; G\'omez et al. in prep.).

\subsection{Distribution on the sky}

Figure \ref{fig:sky_haloA} shows the distribution in the sky of the RGB
stars present in the stellar halo of \Aq{A}. The different panels in
Fig.~\ref{fig:sky_haloA} correspond to stars located at increasing
distance from the Sun, which is assumed to be at (-8, 0, 0) kpc
with respect to the halo center (for this strongly prolate halo, the
major axis is very nearly aligned with the $z$-axis of the simulation
reference frame). Particles that have the same color represent stars
formed in the same parent galaxy.

This figure shows that the distribution of stars in the accreted halo 
is very smooth in the inner 10 kpc. Substructure becomes
apparent at $\sim 20$~kpc and dominates the halo beyond $\sim$30
kpc. The substructures are often diffuse, particularly at small
distances from the galactic center. This is because their progenitor
satellites are relatively massive. For example, the most prominent
streams in Fig.~\ref{fig:sky_haloA} are those in magenta (visible at
all distances), green (dominant in the very center), blue-gray (which we
describe below as a Sgr analogue) and light green (prominent beyond 30
kpc). These contribute $1.3 \times 10^8$, $1.4 \times 10^8$, $2.2 \times
10^7$ and $4.5 \times 10^7$ M$_\odot$ respectively, i.e. all together
they add up to 85\% of the stellar mass in the halo of \Aq{A}.

The distribution of substructure on the sky is anisotropic, and
appears to be preferentially found along a ``ring''. In this region, a
system of streams very similar to those of the Sagittarius dwarf
galaxy is apparent (shown in blue-gray).  Even the central regions of
the stellar halo of \Aq{A} are not isotropic on the sky -- a bar-like
feature is visible towards the galactic center. The central region of
this stellar halo is triaxial, which may also be true in the case of
the Milky Way \citep{newberg2007}. However, it is likely that the
degree of triaxiality would decrease if a massive stellar disk were
included in the simulation \citep[see e.g.][]{Dubinski,Debattista,SK,SK10,tissera}.

Figure \ref{fig:sky_halo_all} shows the sky distribution of stars
located between 10 and 30~kpc (top) and 30 to 50~kpc (bottom) for the
remaining halos. Substructure is apparent in all cases, but there are
large variations in its coherence, surface brightness and distribution
across the sky. As in the case of \Aq{A} , the most significant
overdensities are found at 10 - 30 kpc, although abundant substructure
is also present at larger distances.
 
\begin{figure}
\begin{center}
\includegraphics[height=0.2\textheight]{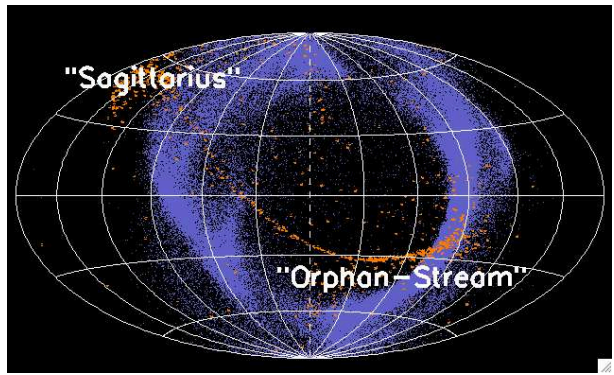}
\end{center}
\caption{Sky distribution of RGB stars in \Aq{A} from the 
``Sagittarius'' (blue) and ``Orphan'' (orange) streams
  analogues. Only those stars located at distances smaller than 80 kpc
  from the ``Sun'' are shown.
  \label{fig:sgr}}
\end{figure}

The object found in \Aq{A} that resembles Sagittarius and its streams
became a satellite at redshift 1.74 (9.7 Gyr ago), and had a total
mass of $2.9 \times 10^{10} \sm$ and a stellar mass of $2.7 \times
10^{7} \sm$ at the time of infall. A bound core survived until the
present day with approximately 25\% of the initial stellar
mass. Fig.~\ref{fig:sgr} shows that its streams cover a similar area
of the sky and have distances comparable to those of the Sagittarius
streams in the Milky Way halo.

We also find structures that resemble the Orphan stream. Although they
are not common at small radii, very low surface brightness thin
streams are present, and can be found as close as 10 kpc from the halo
center. Typically they have elongated orbits, and their characteristic
apocentric distances range from 30~kpc up to $\sim 130$~kpc. This
explains why they have remained coherent despite the triaxial shape of
the halo (and the graininess of the potential). Such streams originate
in low mass galaxies. For example, the progenitor of the thin stream
shown in Fig.~\ref{fig:sgr} had a total mass of $1.2 \times 10^8 \sm$,
a stellar mass of $10^5 \sm$ and was accreted into \Aq{A} at $z = 
5$, i.e.\ $\sim 12.5$~Gyr ago.

\begin{figure}
\begin{center}
\includegraphics[height=0.17\textheight]{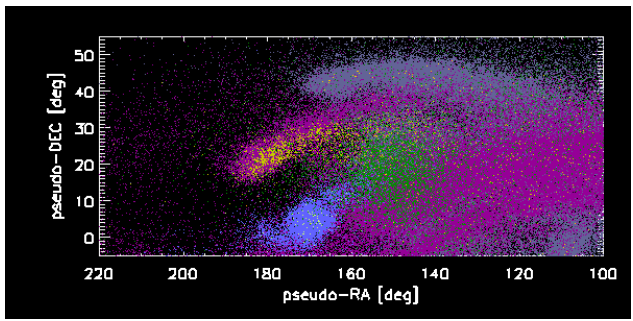}
\end{center}
\caption{Distribution of MSTO stars in the Aquarius ``Field of
Streams'' colored according to provenance. The region of the sky 
shown here has a similar extent to the SDSS footprint, but the model ``stars'' are located in a thin
slice of 1 kpc width at a distance of $\sim 35$ kpc.  Streams from
different progenitors overlap on the sky because of the correlated 
infall directions and also due to group infall (both characteristic of
$\Lambda$CDM). This constitutes a plausible (alternative) 
explanation of the observed bifurcation of the Sgr stream.
\label{fig:fos}}
\end{figure}
 
\subsection{The ``Field of Streams''}

The previous figures show that in our simulations, streams from
different progenitors frequently overlap on the sky. This is due
mostly to the correlated infall pattern of the progenitor satellites
from which these streams originate. For example, \Aq{A} remains
embedded in the same large-scale coherent filament for $\sim10$~Gyr
before the present day, and many of its satellites have formed in this
structure \citep[cf][also Lovell et al. 2010, Vera-Ciro et al. in prep.]{libeskind}.

This is exemplified in Fig.~\ref{fig:fos}, which plots the
distribution of MSTO stars (generated following our simple
prescription) for \Aq{A} in a thin distance slice through a region
similar to the SDSS footprint known as the ``Field of Streams''
\citep{vasily-fos}. Here different colors indicate different
progenitor systems. The characteristics of this Aquarius ``Field of
Streams'' are similar to those observed in SDSS. In particular, we see
streams of stars that follow similar paths on the sky, resembling the
bifurcation discovered by \citet{F06}, as well as various broad
overdensities. Note that some of the bifurcations that are apparent do
not necessarily arise from the overlap of streams on the same orbit
with different orbital phase, but instead correspond to the overlap of
streams of different origin. This implies that measurements of
position and distance alone may be insufficient to associate
overdensities in nearby regions of the sky with a single parent
object; therefore some caution is required when such associations are
used to constrain the shape of the underlying gravitational potential.

\subsection{Quantifying the amount of substructure}

Following \cite{bell}, we have selected stars from SDSS-DR7 with
$140\ndeg \le \alpha \le 220\ndeg$ and $0\ndeg \le \delta \le
60\ndeg$, $0.2 < (g$--$r) < 0.4$  (characteristic of
the halo MSTO), and with apparent magnitude $18.5 \le
r \le 22.5$. If an absolute magnitude $M_r \sim 4.5$ is assumed,
these stars would probe a distance range of 7 up to 35~kpc. In
practice, this simple `tomography' of the halo is subject to
various uncertainties. 

For example, using data from SDSS Stripe 82, \citet{juric}  estimated
that at the bright end, 5\% of the point-like sources with $0.2 <
(g$--$r) < 0.3$ are QSOs. At $r \sim 22$, this fraction increases to
34\%, which extrapolated to $r\sim22.5$, implies a non-negligible
contamination of 50\%.  Another complication arises from photometric
errors. Although these are relatively small, the color
error at the faint end is $\sigma_{g-r}\sim 0.2$, i.e. comparable to
the width of the MSTO color selection box
\citep{ivezic-arxiv}. Therefore, a large number of lower main sequence
stars are scattered into the MSTO region, leading to an incorrect
assignment of absolute magnitudes. We consider both of these effects in
our simulations.

\begin{figure}
\begin{center}
\includegraphics[height=0.35\textheight]{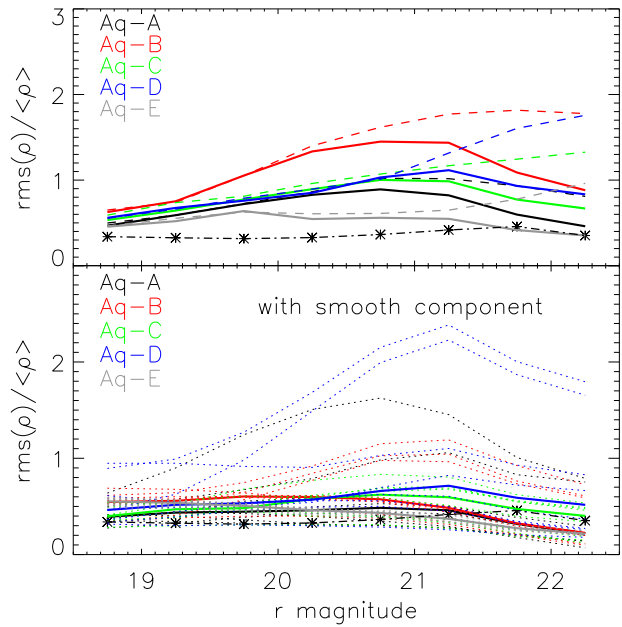}
\end{center}
\caption{ {\it RMS} to mean density of MSTO ``stars'' in a region of
  the sky similar in location and extent to the SDSS footprint as a
  function of apparent magnitude for our accreted Aq-halos.  {\it Top
    panel:} The dashed curves indicate the raw measurements, while the
  solid curves take into account photometric errors and QSO
  contamination.  {\it Bottom panel:} A smooth component with 10\% of the
  total stellar halo mass has been added to each Aq-halo. The
  measurements for eight observers located along the ``solar'' circle
  are given by the dotted curves, while their median value is shown in
  solid. In both panels, the black asterisks correspond to the values estimated
  from SDSS data for the Milky Way.  \label{fig:fraction_sdss}}
\end{figure}
 
We measure the {\it RMS} in the sky density of sources identified as
halo MSTO stars in the SDSS as a function of apparent magnitude, in
cells of 2$\times$2 deg$^2$. To compare with the simulations, we
derive the apparent magnitude of each model MSTO stellar particle as
$m_r = 5 \log d - 5 + M_r + \Delta M_r$. Here $d$ is the distance
to the observer, and $M_r$ is a random variable following a Gaussian
distribution with $\langle M_r \rangle = 4.5$ and $\sigma = 0.9$ (to
mimic the spread in the absolute magnitude of the MSTO). The term $\Delta M_r$
accounts for the uncertainty due the color
error $\sigma_{g-r}$, and is computed as $\Delta M_r =
{\rm d} M_r/{\rm d}(g-r) \times \sigma_{g-r} \times u$, where ${\rm d} M_r/{\rm
  d}(g-r)$ is the slope of the main sequence track \citep[derived
from the photometric parallax relations by][]{juric}, and $u$ is a
normally distributed random variable with zero mean and unity dispersion.
 
In our simulations, we place eight observers along a circle of 8~kpc
radius from the halo center, in a plane perpendicular to the minor
axis of the dark matter halo (an approximation to the Solar
circle). We select those MSTO ``stars'' located in the same region of
the sky and with the same apparent magnitude range as the SDSS sample,
and we measure the (Poisson-corrected) {\it RMS} in the projected
surface density of these stars for each of our halos. Finally,
we add the expected contamination by QSOs as described above.  In this
way, we obtain the fractional {\it RMS} which may be compared to the
measurement from SDSS.

In the top panel of Fig.~\ref{fig:fraction_sdss} the solid curves denote
the median (of each of the 8 independent observers) for each halo.
The dashed curves correspond to the raw {\it RMS} obtained without taking
into account the effect of photometric errors and QSO
contamination. This figure shows that our halos tend to have
systematically larger fractional {\it RMS} values than found in the
SDSS data. 
The bottom panel of Fig.~\ref{fig:fraction_sdss} shows the {\it RMS}
statistic when we include the contribution of a smooth component. We
plot both the median {\it RMS} for each halo (solid) as well as that
measured by
each of the 8 observers (dotted). Here we have assumed 10\% of the
total stellar halo mass to be distributed following a Hernquist
profile with scale radius of 1.25 kpc, although our results are not
strongly dependent on the exact value of this parameter (note that at
$r \sim 19$, its contribution is $\sim 30$\%). This component would
represent stars formed in the Galaxy itself, which, by definition, are
absent in the C10 models of accreted halos. The main effect is to
lower the contrast between cells with and without overdensities.  The
comparison to the SDSS results improves noticeably, indicating that
there is indeed room for an in-situ component to be present in the
stellar halo of the Milky Way \citep[see also][]{zolotov}.

\section{Summary}

We have discussed the substructure present in the stellar halos of
Cooper et al. (2010) which were formed by the disruption of accreted
satellites within the Milky Way mass $\Lambda$CDM halos of the
Aquarius project. Diffuse features such as the Pisces and Virgo
overdensities and the Hercules-Aquila cloud are relatively common, as
are narrow, faint ``orphan'' streams. One of our halos even contains a
structure similar to the Sagittarius Stream. We have found that
substructures are not distributed isotropically on the sky. This is
likely to be related to the correlated infall of satellites accreted
along large-scale filaments, a characteristic feature of the cold dark
matter model \citep{libeskind,li-ah,lux}. As a consequence, chance
alignments between streams from different progenitors may occur (as in
our ``Sagittarius'' example), which may be easily misinterpreted as
streams from the same object with different orbital phase. We find
that the fraction of substructure in our models is somewhat higher
than is estimated for the Milky Way.  The addition of a smooth
spheroidal component that contains 10\% of the total stellar halo mass
is required to match the simulations the SDSS data. This might suggest
that a fraction of the halo stars formed in-situ. However, the
discrepancy could also, however, be the result of limitations in our
modeling techniques or uncertainties in the interpretation of the SDSS
data.

\acknowledgements

We are especially indebted to Volker Springel. AH acknowledges funding
support from the European Research Council under ERC-StG grant
GALACTICA-240271. APC acknowledges an STFC postgraduate studentship,
SMC the support of a Leverhulme Research Fellowship and CSF
acknowledges a Royal Society Wolfson Research Merit Award.

\end{document}